\documentclass{elsart5p}

 \usepackage{graphicx}

\usepackage{amssymb}

\begin{document}

\begin{frontmatter}



\title{Spin chirality and electric polarization in multiferroic 
compounds\\ $R$Mn$_2$O$_5$ ($R=$Ho, Er)}


\author[label1]{Shuichi Wakimoto}
\author[label2]{Hiroyuki Kimura}
\author[label2]{Mamoru Fukunaga}
\author[label2]{Keisuke Nishihata}
\author[label1]{Masayasu Takeda}
\author[label1]{Kazuhisa Kakurai}
\author[label2]{Yukio Noda}
\author[label3]{Yoshinori Tokura}

\address[label1]{Quantum Beam Science Directorate, Japan Atomic Energy Agency,
Tokai, Ibaraki 319-1195, Japan}
\address[label2]{Institute of Multidisciplinary Research for Advanced Materials, Tohoku University, Katahira, Sendai 980-8577, Japan}
\address[label3]{Department of Applied Physics, University of Tokyo, Tokyo 113-8656, Japan}

\begin{abstract}

Polarized neutron diffraction experiments have been performed on multiferroic materials $R$Mn$_{2}$O$_{5}$ ($R=$Ho, Er) under electric fields in the ferroelectric commensurate (CM) and the low-temperature incommensurate (LT-ICM) phases, where the former has the highest electric polarization and the latter has reduced polarization.  It is found that, after cooling in electric fields down to the CM phase, the magnetic chirality is proportional to the electric polarization.  Also we confirmed that the magnetic chirality can be switched by the polarity of the electric polarization in both the CM and LT-ICM phases.  
These facts suggest an intimate coupling between the magnetic chirality and the electric polarization.  
However, upon the transition from the CM to LT-ICM phase, the reduction of the electric polarization is not accompanied by any reduction of the magnetic chirality, implying that the CM and LT-ICM phases contain different mechanisms of the magnetoelectric coupling. 

\end{abstract}

\end{frontmatter}

\section{Introduction}
\label{}

Multiferroic materials $R$Mn$_2$O$_5$ ($R$: rare earth) show antiferromagnetic order and ferroelectricity concomitantly at low temperatures with strong coupling, resulting in a remarkable magnetoelectric effect~\cite{TKimura_03,Hur_04}.  The understanding of the coupling mechanism is one of the central issues.

Generally $R$Mn$_2$O$_5$ shows antiferromagnetic order with incommensurate (ICM) modulation at $\sim 45$~K.  At lower temperature, typically $\sim 35$~K, the magnetic structure transforms into commensurate (CM) with propagation vector ${\bf q} = (0.5, 0, 0.25)$ and electric polarization appears concomitantly~\cite{Fukunaga_07}.  Then at lower temperature, $<\sim 10$~K, the magnetic structure becomes ICM again, and the electric polarization becomes suppressed.
In the case of $R=$Ho, however, the electric polarization in the low-temperature ICM (LT-ICM) recovers by application of magnetic fields~\cite{Higashiyama_05}.  Later, it has been reported that the recovery of the electric polarization coincides with a transition of magnetic structure from LT-ICM to CM~\cite{Kimura_06}.
%
%
In addition, it is shown that the CM magnetic structure contains spirals of the Mn spins with magnetic unit cell of $(2a, b, 4c)$~\cite{Kimura_07}.  These facts imply that the magnetic spiral with CM modulation periodicity is preferable for the ferroelectric phase.  On the other hand, the LT-ICM phase is less understood due to the difficulties in magnetic structural analysis.

\begin{figure*}
\includegraphics[width=7.2in]{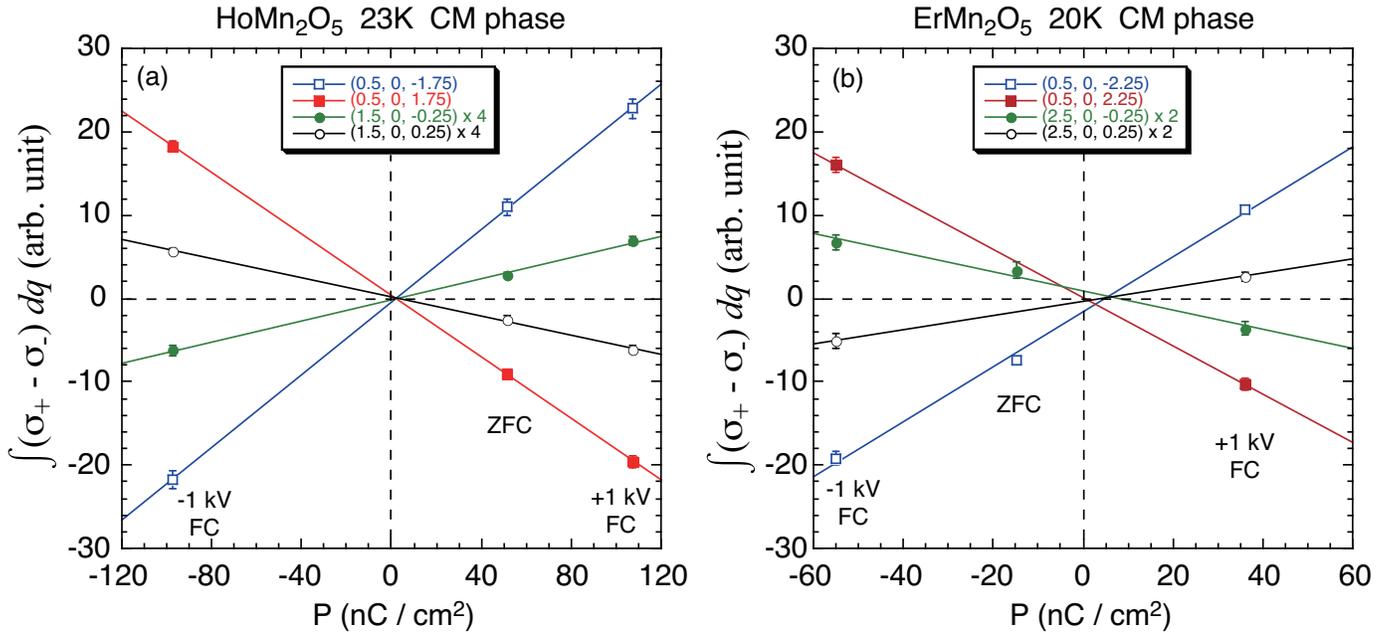}
\caption{
Chiral terms for (a) HoMn$_2$O$_5$ and (b) ErMn$_2$O$_5$ as a function of the electric polarization along the $b$-axis.  The vertical axis is $\int(\sigma_{+}-\sigma_{-})dq$, which is integrated one dimensionally along the $a^{*}$-direction, representing the chirality.  For HoMn$_2$O$5$, the chirality is measured at $(1.5, 0, 0.25)$ (open circles), $(1.5, 0, -0.25)$ (filled circles), $(0.5, 0, 1.75)$ (filled squares), and $(0.5, 0, -1.75)$ (open squares).   For ErMn$_2$O$_5$, it is measured at $(2.5, 0, 0.25)$ (open circles), $(2.5, 0, -0.25)$ (filled circles), $(0.5, 0, 2.25)$ (open squares), and $(0.5, 0, -2.25)$ (filled squares).
}
\end{figure*}

To date, two distinct models have been suggested as an origin of the magnetoelectric coupling in the Mn multiferroics.  One is an inverse Dzialoshinski-Moriya (DM) interaction model~\cite{Katsura_04} and the other is called exchange striction model~\cite{Chapon_06}.  The former produces polarization from a cross term of spins ${\bf S}_i \times {\bf S}_j$, while the latter based on ${\bf S}_i \cdot {\bf S}_j$ products.
The former model has been examined using the related compound TbMnO$_3$ by polarized neutron diffraction with electric field and found that the magnetic chirality ${\bf S}_i \times {\bf S}_j$ (in another word, clockwise or counter clockwise) can be indeed controlled by the polarity of the electric field~\cite{Yamasaki_07}.  However, it has not been tested for the $R$Mn$_2$O$_5$ family.

Here we report results of polarized neutron diffraction experiments under electric field in the CM and LT-ICM phases of HoMn$_2$O$_5$ and ErMn$_2$O$_5$ performed to understand the role of the magnetic chirality to the magnetoelectric coupling in these compounds.  Results show that the magnetic chirality in the CM phase can be switched by the application of electric field and, after the field-cooling process, the electric polarization is proportional to the magnetic chirality, which apparently supports the ${\bf S}_i \times {\bf S}_j$ model.  However, when the system transforms into the LT-ICM phase on cooling with electric field, the magnetic chirality merely changes little while the electric polarization decreases drastically.  This may imply that the CM and the LT-ICM phases contain different mechanisms of the magnetoelectric coupling.

\section{Experimental procedure}

Polarized neutron diffraction experiments were done at the TAS-1 spectrometer installed at the JRR-3M reactor in JAEA.  A combination of a pyrolytic graphite (PG) monochromator and a double-focusing Heusler analyzer has been used with a spin flipper in front of the analyzer.  A guide field around the sample was kept either parallel to the momentum transfer {\bf Q} (horizontal field) or vertical by a Helmholtz coil.  The incident neutron energy $E_i = 14.7$~meV and the collimator sequence of 40$'$-80$'$-80$'$-open were used.  PG and sapphire filters were located in front of the sample to eliminate higher order and fast neutrons, respectively.  In this configuration, we can analyze the polarization of the diffracted neutrons where the incident beam is unpolarized.  

Disk-shaped single crystals of $R$Mn$_2$O$_5$ ($R=$Ho, Er) with a thickness of $\sim 1$~mm along the $b$-axis were placed in a closed-cycle He-gas refrigerator.  The $b$-axis was set vertical, so that one can access to $(H, 0, L)$ type reflection.  Electrodes were attached on gold masks on the $b$-surface.  
In the measurements, we cooled the sample from 65~K under electric field up to 1~kV to a target temperature, 20 (23)~K for $R=$Er (Ho) for the CM phase measurements and    4~K for the LT-ICM phase measurements.  Then we perform polarized neutron diffraction followed by a warming procedure without electric field.  Pyrocurrent has been measured in the warming process up to the 55K, well above the ferroelectric transition temperature, to evaluate the electric polarization that the system holds in the polarized neutron measurements.  

\begin{figure*}
\includegraphics[width=6in]{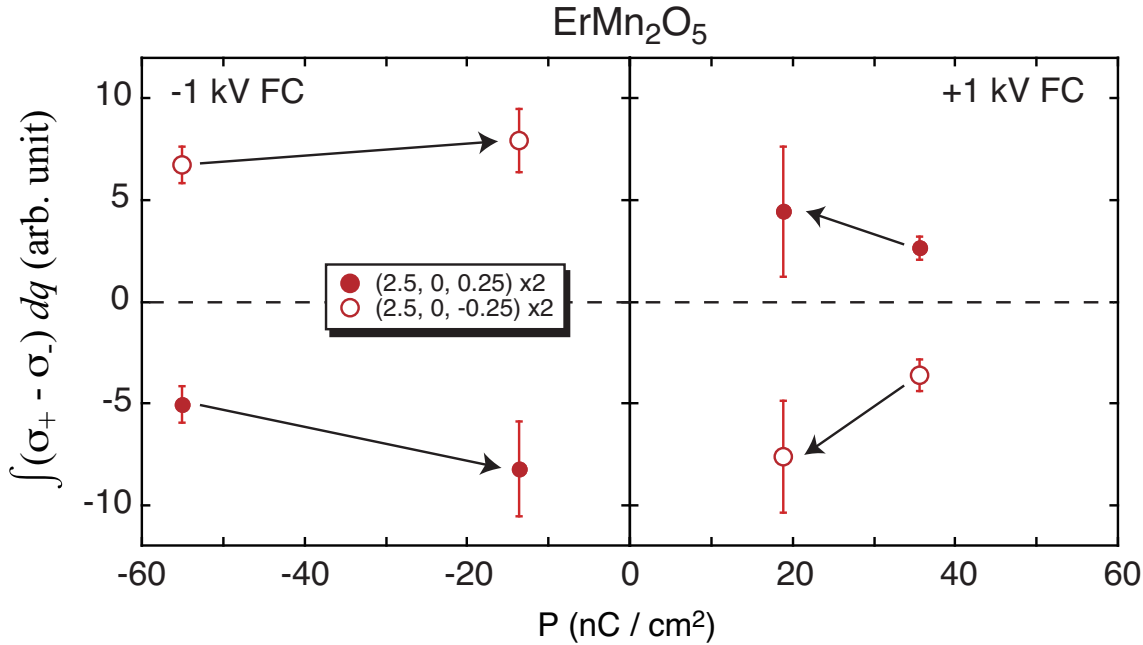}
\caption{
Variation of magnetic chirality and the electric polarization along the $b$-axis for ErMn$_2$O$_5$ between the CM phase (20~K) and the LT-ICM phase (4~K).  The magnetic chirality has been measured at $(2.5, 0, 0.25)$ (closed circles) and $(2.5, 0, -0.25)$ (open circles).  (Note that the exact $q$-position in the LT-ICM phase is $(2.5, 0, \pm 0.27)$. )  The arrows express the direction of change from the CM to LT-ICM phase.
}
\end{figure*}

\section{Polarized neutron diffraction}

Before we present the results, we briefly review how the magnetic chirality can be detected by the polarized neutron diffraction.
The polarization of the incident and diffracted neutrons, ${\bf P_i}$ and ${\bf P_f}$, are connected by the following formula in the magnetic scattering cross section from a magnetic structure with chirality~\cite{Blume_63}:
\begin{eqnarray}
{\bf P_f} \sigma &=&  \{ -{\bf P_i} + 2({\bf P_i}\cdot\hat{\bf M})\cdot\hat{\bf M} \} \cdot \sigma_m 
						+ i({\bf M}\times{\bf M}^{*}),
\end{eqnarray}
where 
\begin{eqnarray}
{\bf M} &=& \sum_{j}{{\bf S}_{j\perp}e^{i{\bf Q}\cdot{\bf r}_j}}\\
             &=& \sum_{j}{{\bf S}_{j\perp}\cos{{\bf Q}\cdot{\bf r}_j}}
                      + i \sum_{j}{{\bf S}_{j\perp}\sin{{\bf Q}\cdot{\bf r}_j}}.
\end{eqnarray}
Here,$\sigma$ is the total neutron cross-section, $\sigma_m$ is the magnetic cross-section, and ${\bf S}_{j\perp}$ is a component of the $j-$th spin perpendicular to the scattering vector {\bf Q}.  The first term in Eq. (1) means that the neutron polarization makes $\pi$-precession with respect to {\bf M}.  The second term expresses the chiral structure contribution indicating that the final polarization is forced to align along the chiral vector $i({\bf M}\times{\bf M}^{*}) // {\bf Q}$.  
In our experimental set-up, the incident neutron is unpolarized, $P_i=0$, which makes the contribution of the first term zero.  Therefore, the difference in the spin (+) and (-) of the diffracted beam, ($\sigma_{+}-\sigma_{-}) \propto P_f\sigma$, in the horizontal field mode observes the magnetic chirality directly.

In this paper, we use a quantity $\int(\sigma_{+}-\sigma_{-})dq$ which is integrated one dimensionally along the $a^{*}$-direction in order to express the magnetic chirality.

\section{Results}

\subsection{CM phase}

Since the CM phase has the most pronounced ferroelectric polarization, we have started with the measurements of the CM phase.

Figures 1 (a) and (b) show magnetic chirality in the CM phase of HoMn$_2$O$_5$ and ErMn$_2$O$_5$, respectively, as a function of the electric polarization along the $b$-axis.  The data at $\sim 50$~nC/cm$^2$ for Fig. 1 (a) and those for $\sim -15$~nC/cm$^2$ for Fig. 1 (b) are measured after cooling under a zero field, while the others are either data measured after cooling in $\pm 1$~kV.   
We note that our samples show spontaneous polarization after the zero-field-cooling.  This is because of a sample-dependent peculiarity that polarized domains are imbalanced even after the zero-field-cooling process~\cite{Fukunaga_07_K}.
The chirality has been measured at both Q-positions near the $a^{*}$-axis, $(1.5, 0, \pm 0.25)$ for HoMn$_2$O$_5$ and $(2.5, 0, \pm0.25)$ for ErMn$_2$O$_5$, and near the $c^{*}$-axis, $(0.5, 0, \pm 1.75)$ for HoMn$_2$O$_5$ and $(0.5, 0, \pm2.25)$ for ErMn$_2$O$_5$.  Since the neutrons see magnetic component that is perpendicular to {\bf Q}, the former corresponds to the magnetic chirality in the $bc$-plane while the latter in the $ab$-plane.
For both samples, particularly for HoMn$_2$O$_5$, it is demonstrated that the chirality is proportional to the electric polarization even with the zero-field-cooling data.

\subsection{LT-ICM phase}

It is also an important question to ask if the magnetic chirality and the electric polarization change accordingly at the transition from the CM phase to the LT-ICM phase where the electric polarization is known to be suppressed.

It should be noted here that the LT-ICM phases of HoMn$_2$O$_5$ and ErMn$_2$O$_5$ are different.  The former is two-dimensionally incommensurate; that is, the modulation vector is $(q_x, 0, q_z)$ where $q_x \sim 0.48$ and $q_z \sim 0.27$.  In contrast, the latter is one-dimensionally incommensurate; the modulation vector is $(0.5, 0, q_z)$ where $q_z \sim 0.27$.  Thus, at the CM to LT-ICM transition, the magnetic peak splits in the $a^*$-direction for HoMn$_2$O$_5$, while not for ErMn$_2$O$_5$.
We have observed that the split magnetic peaks in the LT-ICM phase of HoMn$_2$O$_5$ show opposite sign of $(\sigma_{+}-\sigma_{-})$ to each other~\cite{waki_up}.  It is somewhat ambiguous to discuss the magnetic chirality from such the altered pair of the LT-ICM peaks of HoMn$_2$O$_5$ without the detailed magnetic structure of the LT-ICM phase which has not been solved yet.  However we can avoid such ambiguity in the case of ErMn$_2$O$_5$ since the magnetic chirality can be uniquely determined from the LT-ICM peak which does not split.

Figure 2 summarizes the change of the magnetic chirality measured at $(2.5, 0, \pm 0.25)$ and the electric polarization along the $b$-axis between the CM phase (20~K) and the LT-ICM phase (4~K).  The arrows express the direction of change from the CM to LT-ICM phase.
Remarkably, it is shown that the suppression of the electric polarization in the LT-ICM phase is not accompanied by any reduction of the magnetic chirality, which even increases in the LT-ICM phase.  However, at least it is shown that the magnetic chirality is switched by the change of the polarity of the electric field.

\section{Discussion and summary}

We have shown, by the polarized neutron diffraction studies on the multiferroic compounds $R$Mn$_2$O$_5$ ($R=$Ho, Er) under electric field, that the magnetic chirality of the Mn-spins, i.e., clockwise or counter-clockwise, can be switched by the polarity of the external electric field in the cooling process, both for the CM and LT-ICM phases.  Together with the control of the magnetic chirality of the ferroelectric phase of TbMnO$_3$ by electric field reported by Yamasaki {\it et al.}~\cite{Yamasaki_07}, the present results imply the universality of coupling between electric polarization and magnetic chirality.  
Such coupling is reasonable on the basis of the symmetry consideration.  However the linear relation in Fig. 1 suggests a coupling in the level of microscopic mechanism.

There remains an important question for the $R$Mn$_2$O$_5$ system: why the highest electric polarization achieves in the CM structure?  As shown in Fig. 2, the suppression of the electric polarization in the LT-ICM phase cannot be accounted for by the magnetic chirality alone.  A possibility is, on the basis of the inverse DM model, that the CM and LT-ICM phases have different coupling constants between the electric polarization and ${\bf S}_i \times {\bf S}_j$. 
Another possibility is that the CM and LT-ICM phases contain different origins of ferroelectricity: for example, the electric polarization in the CM phase originates from the ${\bf S}_i \cdot {\bf S}_j$ product in addition to the ${\bf S}_i \times {\bf S}_j$ product.  The reduction of the electric polarization in the LT-ICM phase might be caused by suppression of ${\bf S}_i \cdot {\bf S}_j$ product due to magnetic transition.
Detailed analyses of both crystal and magnetic structures of the LT-ICM phase is necessary to address this question.

In summary, we have performed polarized neutron diffraction study under electric field using $R$Mn$_2$O$_5$ ($R=$Ho, Er).  We observed a linear relation between the magnetic chirality of Mn spins and the electric polarization in the ferroelectric CM phase.  Also we confirmed that the magnetic chirality can be switched by the polarity of electric field in both the CM and LT-ICM phases.  However, upon cooling from the CM to the LT-ICM phase, magnetic chirality shows small increase while the electric polarization decreases dramatically.  This suggests different origins of the magnetoelectric coupling in the CM and LT-ICM phases.

\vspace{5mm}
\noindent
{\bf Acknowledgments}

Authors thank Kay Kohn for invaluable discussion.  This work was supported by Grant-In-Aid for Scientific Research on Priority Areas "Novel State of Matter Induced by Frustration" (No. 19052001 and No. 19052004) and by a Grant-in-Aid for Scientific Research (B) No. 16340096 from the Ministry of Education, Culture, Sports, Science and Technology, Japan.
%

%




\end{document}